# Ferromagnetic Domain Distribution in Thin Films During Magnetization Reversal

W.-T. Lee [1], S. G. E. te Velthuis [2], G. P. Felcher [2], F. Klose [1], T. Gredig [3], and E. D. Dahlberg [3].

[1] Spallation Neutron Source, Oak Ridge National Laboratory
[2] Material Science Division, Argonne National Laboratory
[3] Department of Physics, University of Minnesota

Abstract

We have shown that polarized neutron reflectometry can determine in a model-free way not only the mean magnetization of a ferromagnetic thin film at any point of a hysteresis cycle, but also the mean square dispersion of the magnetization vectors of its lateral domains. This technique is applied to elucidate the mechanism of the magnetization reversal of an exchange-biased Co/CoO bilayer. The reversal process above the blocking temperature $T_b$ is governed by uniaxial domain switching, while below $T_b$ the reversal of magnetization for the trained sample takes place with substantial domain rotation.

Polarized neutron reflectometry (PNR) was introduced in the beginning of the '80s to map the magnetic profiles of thin films and multilayers [1-3]. This technique has been applied to study systems in which the magnetic structure consists of a stack of laterally uniform magnetic layers. The experiments reveal the depth-dependence of the magnetization vectors caused by magnetic interactions between different layers. Polarized neutron reflectometry can also be instrumental in understanding a different, but still outstanding problem in magnetism: the breakdown into domains of a ferromagnet during the hysteresis cycle [4]. The neutron reflectivity studies are limited to samples in the form of thin flat films. This is the form of many magnetic systems created recently for diverse applications, from magnetic recording to magnetic memory to sensors. Magnetic domains significantly impact the performance of these devices. In most applications the ferromagnetic layers are so thin, that only a single magnetic domain can be energetically stable through the thickness. While passing through a hysteresis loop, however, the film may break down into a collection of *lateral* magnetic domains, each characterized by a size and a direction of magnetization. In spite of the progress made in the use of microscopic and scattering probes, the problem of observing these domains - especially at some distance below the surface - remains formidable. Yet, as demonstrated in this letter, a statistical measure of the domain distribution can be obtained directly from PNR.

In a PNR experiment, the intensity of the neutrons reflected from a surface is measured as a function of the component of the momentum transfer that is perpendicular to the surface, $q_z = 4\pi\sin\theta/\lambda$, where $\theta$ is the angle of incidence (and reflection) and $\lambda$ the neutron wavelength. Since $q_z$ is a variable conjugate of the depth z from the surface of the film, a scan over suitable range of $q_z$ provides excellent information on the chemical and magnetic depth profile of the film. When the incident neutrons are polarized along a applied magnetic field **H**, and the polarization after reflection is analyzed along the same axis, four reflectivities are recorded: $R^{++}$, $R^{+-}$, $R^{-+}$, $R^{--}$. The first (second) sign refers to the incident (reflected) neutron polarization with respect to **H**. In a simplified way it can be said that $R^{++}$ and $R^{--}$ measure the magnetization component along the field, $M$, while the magnetic components perpendicular to the field give rise to the "spin flip" intensities $R^{+-}$ and $R^{-+}$. If the film is laterally homogeneous, a fitting procedure allows a model magnetic profile to be obtained from the spin-dependent reflectivities.

When the direction of magnetization is in the film's plane, the following quantities, expressed in terms of the angle $\varphi$ between **H** and **M**, can be easily determined from the specular reflectivities [5]:

$$\frac{R^{++}(\varphi)-R^{--}(\varphi)}{R_s^{++}(0°)-R_s^{--}(0°)} = \frac{R^{+}(\varphi)-R^{-}(\varphi)}{R_s^{+}(0°)-R_s^{-}(0°)} = \cos\varphi, \quad (1)$$

$$\frac{R^{-+}(\varphi)}{R_s^{-+}(90°)} = \sin^2\varphi. \quad (2)$$

These relations are strictly valid at all values of $q_z$ provided that the direction of **M** is constant along the thickness of the film. The normalizing quantities, $R_s(0°)$ and $R_s(90°)$, refer to reflectivities measured with **M** aligned parallel and perpendicular to the neutron polarization, respectively. The single-superscript reflectivities are those measured without polarization analysis, i.e., $R^+ = R^{++} + R^{+-}$ and $R^- = R^{--} + R^{-+}$ (for this configuration of fields, $R^{+-} = R^{-+}$).

Lateral magnetic domains cause a broadening of the reflected beam. The width of the specular reflection is consistent with a coherence length of the neutron beam estimated to be of the order of tens of microns. In the case of lateral domains smaller than the coherence length, an off-specular diffuse component appears centered around the specular reflection [6-9], and its width is inversely proportional to the magnetic domain size in the absence of correlations. For larger domain sizes, the intensities reflected from different domains superimpose incoherently in the specular beam. The terms in $\varphi$ of Eq. (1) and Eq. (2) must now be interpreted as averages across the sample plane. While the term $<\cos\varphi>$ may be measured as well by conventional magnetometry, $<\sin^2\varphi>$ provides new information leading to the mean square dispersion of the domain orientations $\chi^2$:

$$\chi^2 = \langle\cos^2\varphi\rangle - \langle\cos\varphi\rangle^2 = \left\{1 - \frac{R^{-+}(\varphi)}{R_s^{-+}(90°)}\right\} - \left\{\frac{R^{++}(\varphi)-R^{--}(\varphi)}{R_s^{++}(0°)-R_s^{--}(0°)}\right\}^2 \quad (3)$$

These quantities are constant, and independent of $q_z$, if the domains extend through the entire thickness of the film.

These ideas were applied in the study of the magnetic behavior of a partially oxidized Co film [10, 11], which exhibits exchange bias [12]. A 120 Å thick polycrystalline Co film was deposited on a silicon substrate by magnetron sputtering. Its surface was then oxidized in ambient atmosphere to form a 30 Å thick CoO top layer. Since the Co layer was thinner than a domain wall (~ 500 Å) [4], only one domain was expected along the sample's thickness. At the same time, the shape anisotropy constrained the magnetization to be in the plane of the film. The neutron measurements were carried out at the POSY I reflectometer at the Intense Pulsed Neutron Source of Argonne National Laboratory. Figure 1 shows a typical spin-dependent reflectivity pattern of the film. The reflectivity was taken at 10 K and close to remnant magnetization [11]. The fitted profile of the scattering amplitude densities gives the thickness of the layers, the interface roughness between the ferromagnetic (FM) Co and the antiferromagnetic (AF) CoO layers, and the ferromagnetic contribution from Co. The antiferromagnetism of CoO is not considered in the analysis, because for this range of $q_z$ the scattering properties are averaged over a length scale that well exceeds the antiferromagnetic periodicity of CoO.

Films of this type have been found to have three magnetic phases. At temperatures higher than the Néel temperature $T_N$, the magnetization follows a square hysteresis loop [13]. As the temperature is lowered, the hysteresis loop becomes S-shaped and exhibits a scaling behavior as a function of the coercive field and a characteristic temperature below $T_N$ [13]. If the sample is field-cooled to below the "blocking temperature" $T_b$ [14], exchange bias appears - the hysteresis loop is no longer symmetric with respect to the applied field. Fig. 2 shows the hysteresis loops of our film above and below $T_b$, at 140 K and 10 K respectively, after field cooling in +5 kOe from room temperature. The cooling and measurement fields were along the same axis, parallel to an arbitrary direction in the film surface. At 140 K, the hysteresis loop is symmetric with a coercive field $H_C$ = 100 Oe. At 10 K, the initial reversal after field cooling (A2→ B2) has a sharply squared shape and a large reversal field ($H_a$ = -1.1 kOe). Subsequent loops (through C2, D2, etc.) exhibit a more gradual S-shape, with $H_a$ = +300 Oe at one side and $H_a$ = -500 Oe at the other side, giving a bias field $H_e$ = -100 Oe. The insert of Fig. 2 shows the temperature dependence of the bias field of the trained sample, for which $T_b$ ~ 130 K.

To elucidate the mechanism of reversal of the magnetization, PNR measurements were taken close to the reversal points where $M = 0$, as marked in Fig 2. In addition to the specular beam, a weak off-specular scattering was found when the sample was not in a saturated state [11]. The off-specular scattering has the characteristics of Yoneda wings [15], for domains smaller than ~1000 Å. In contrast, the width of the specular reflection is virtually identical to that obtained at saturation. Domains larger than the resolution function of the instrument account for most of the film's surface; we therefore limit ourselves to consider the behavior of the specular reflection.

The PNR results, as obtained from measurements at both the ascending and descending reversal points, are summarized in Fig. 3. $R^+$ and $R^-$ are not presented explicitly, but only the values of $<\cos \varphi>$, as extracted according to Eq. (1): the reflectivities, measured at the reversal points, were divided by the spin-dependent reflectivity at saturation. The $q_z$ dependence of $\chi^2$ was obtained by processing the spin flip intensities $R^{-+}$ according to Eqs. (2) and (3). In this case it was rather simple to obtain the normalizing reflectivity $R_s^{-+}(90º)$. The sample was field-cooled in $H = 2000$ Oe, then the field was switched off and replaced by a guide field of a few Oersted in a perpendicular direction (the guide field was sufficient to orient the neutron spins, but too weak to affect the magnetization).

The $\chi^2$ values are presented for $0.012$ Å$^{-1} < q_z < 0.030$ Å$^{-1}$. In a white beam reflectometer, different $q_z$ are obtained with different neutron wavelengths, and we chose the region of $\lambda$ where polarizer and analyzer had the best efficiency. The lower boundary of $q_z$, instead, was chosen to be close to the value for critical reflection, because there the spin-dependent reflectivity is nearly proportional to the magnetization averaged over the layer thickness. The assumption made, that the orientation of all domains is uniform over the sample thickness, is fairly justified: for the $q_z$ region presented, $\chi^2$ is only slowly changing.

$\chi^2$ is very dependent on temperature and training. To better understand the results, let us consider the extremal cases for $M = 0$. $\chi^2 = 0$ means that all the moments are oriented perpendicular to **H**, implying that the reversal occurs through magnetic domain rotation. In contrast $\chi^2 = 1$ means that half of the moments are parallel, half antiparallel to the field: the reversal occurs by uniaxial domain switching. As seen in Fig. 3(a), $\chi^2$ at 140 K deviates only slightly from unity: above $T_b$, the magnetization reversal occurs primarily through uniaxial domain switching. Similarly, in Fig 3(b), reversal of the untrained film at 10 K (B2) gives $\chi^2$ close to unity. In contrast, the $\chi^2$ values are much smaller for the trained film at the two reversal points C2 and D2. The pertinent $\chi^2$ values, ranging from 0.50 to 0.65, indicate a breakdown in domains with substantial angular spread of their magnetic orientations (for an isotropic distribution of the domains, $\chi^2 = 0.5$).

The present results substantiate a model proposed [13] for the magnetic behavior of Co/CoO. Above the blocking temperature, the S-shaped hysteresis loop has been interpreted in terms of a modified Ising model. The FM Co layer is comprised of a number of domains for which the direction of magnetization is determined by the applied field. [16]. Their reversal takes place over a finite interval of fields because of a range of coupling strengths with different antiferromagnetic CoO domains [17]. The orientation of the sublattice magnetization of the AF domains is not fully locked by the crystalline anisotropy and the coupling between different AF domains. Below $T_b$, however, the AF domains are stabilized. The orientation of the sublattice magnetization of CoO now strongly influences the direction of the FM domains. Unless a strong external magnetic field is present, the FM domains turn their magnetization in the direction optimizing both the coupling with the AF domains and the Zeeman energy [18, 19], giving rise to the rotation of domains we observed.

The results obtained for $T < T_b$ may be compared with those obtained on a different exchange bias system: ferromagnetic Fe coupled with antiferromagnetic $FeF_2$ [20]. The type of neutron measurements carried out in the two cases is similar: the four spin-dependent reflectivities have been measured close to the $M = 0$ points of the hysteresis loop. The results reflect the inherent difference of the two physical systems. In the polycrystalline Co/CoO bilayer a fairly isotropic ferromagnetic domain distribution of Co implies that the AF domain distribution of CoO is equally isotropic [16]. In $FeF_2$ grown semi-epitaxially on MgO(100), AF domains are formed with their sublattice magnetization rigidly aligned along two perpendicular axes. Consequently the orientation of the ferromagnetic domains of polycrystalline Fe, grown on the top of $FeF_2$, had to be more constrained at the $M = 0$ points of the hysteresis loop, as it was confirmed by a careful comparison of the experimental reflectivities with those calculated for a model structure with appropriate distributions of domains [20].

The analysis of neutron reflectivity data presented here is basically model-free, and the experimental data are reduced to quantities with immediate physical significance. It is shown that reflectivity measurements on a thin film composed of a collage of magnetic domains provide not only the average magnetization, but also the mean square dispersion of the domain orientations. Against the danger of oversimplification, several tests were applied. The size of the domains, in a region comprised between 1000 Å and 30 μm, can be monitored from the width of the reflected line. A strongly $q_z$ dependence of $\chi^2$ indicates that direction of magnetization is not uniform throughout the thickness. To cope with such event would require an expansion of Eqs. (1)-(3). Applying this analysis to the $M = 0$ points of the hysteresis cycle in an exchange biased Co/CoO sample revealed different reversal mechanisms above and below the blocking temperature.

If used as described above, PNR does not need detail structural modeling to obtain $<\cos \varphi>$ and $\chi^2$. However these two quantities, when measured at $M = 0$, do not provide a full description of the domain distribution. More detailed and quantitative information could be obtained by measuring the evolution of the magnetic domains as the system is driven along the entire hysteresis cycle. Technically this will become feasible at future spallation neutron sources [21], where it will be possible to measure a reflectivity curve in a matter of minutes, a data acquisition rate comparable to that of conventional magnetization measurements. Both $<\cos \varphi>$ and $\chi^2$ measured by PNR can then be compared with the results of micromagnetic calculations [22]. Even more desirable is the development of a theoretical framework along the lines of the work done to extract information from the magnetization - as obtained by passing with minor hysteresis loops through different paths of metastable states. Starting from the simplest Stoner-Wohlfarth model [23], and continuing through Preisach models of increasing complexity [24], a great insight has been obtained in magnetic systems and at the same time easy recipes have been devised (for instance, the Henkel plots) to characterize them. All this has been possible just using one observable: the average magnetization. If also $\chi^2$, the mean square dispersion of the domain orientations, is available, how much easier or more realistic becomes the analysis of the magnetization process?

*Acknowledgements*

Work done at Argonne National Laboratory was supported by US DOE, Office of Science contract #W-31-109-ENG-38 and by Oak Ridge National Laboratory, managed for the U.S. D.O.E. by UT-Battelle, LLC under Contract No. DE-AC05-00OR22725. Work done at the University of Minnesota was supported by the NSF MRSEC NSF/DMR – 9809364.

FIGURE CAPTIONS

Fig. 1. Polarized neutron reflectivity measured (points) and calculated (lines) for incident neutrons polarized parallel ($R^+$) and antiparallel ($R^-$) to the applied field, at remnant state of the trained film with a guide field of –65 Oe. [11] In the insert shows the scattering length density profile calculated for the two spin states.

Fig. 2. Hysteresis curves (a) above $T_b$ at 140 K and (b) below $T_b$ at 10 K. The labels (A2 at $M$ = saturation, others at $M$ = 0) indicate locations where neutron reflectivities are measured. The insert shows the temperature dependence of the bias field, $|H_e(T)|$.

Fig. 3. Mean square dispersion of lateral domain orientation $\chi^2$ (a) at 140 K, at the reversal points (C1 and D1); (b) at 10K, at reversal points of the untrained film (B2) and the trained film (C2 and D2).

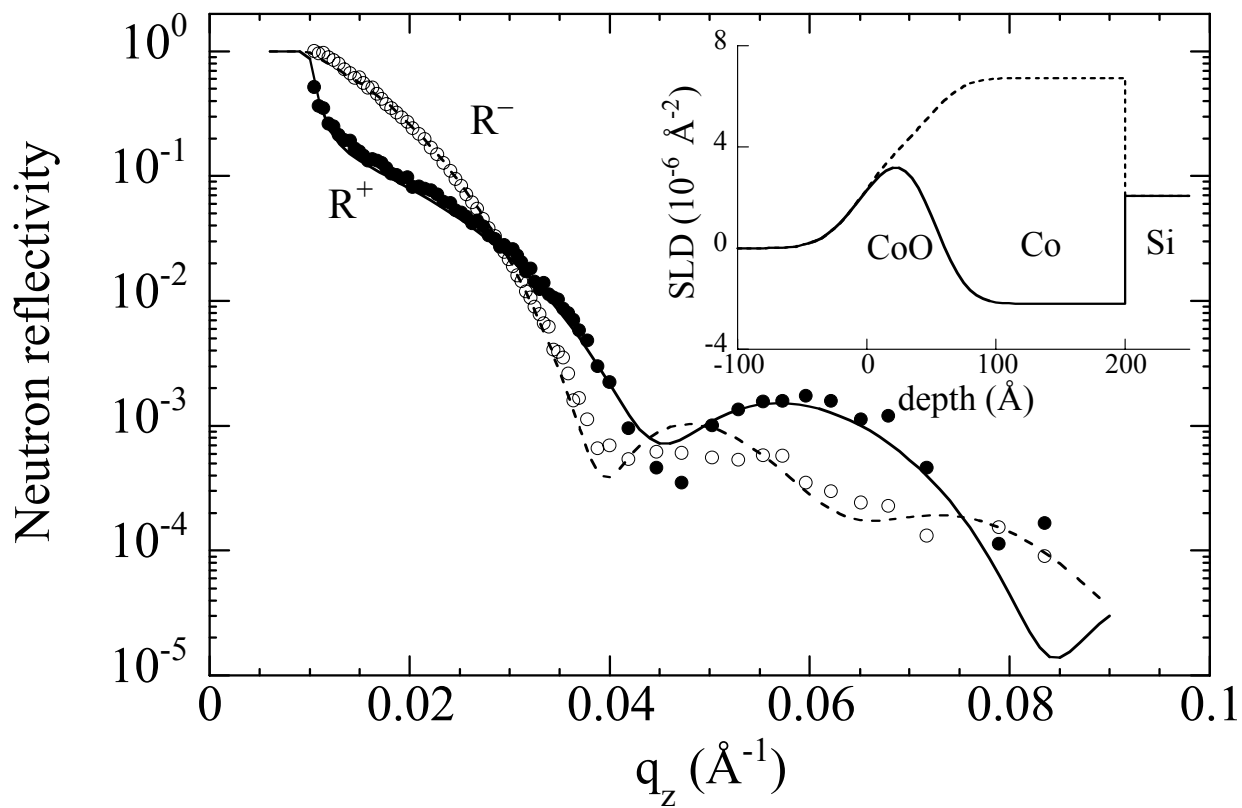

Figure 1

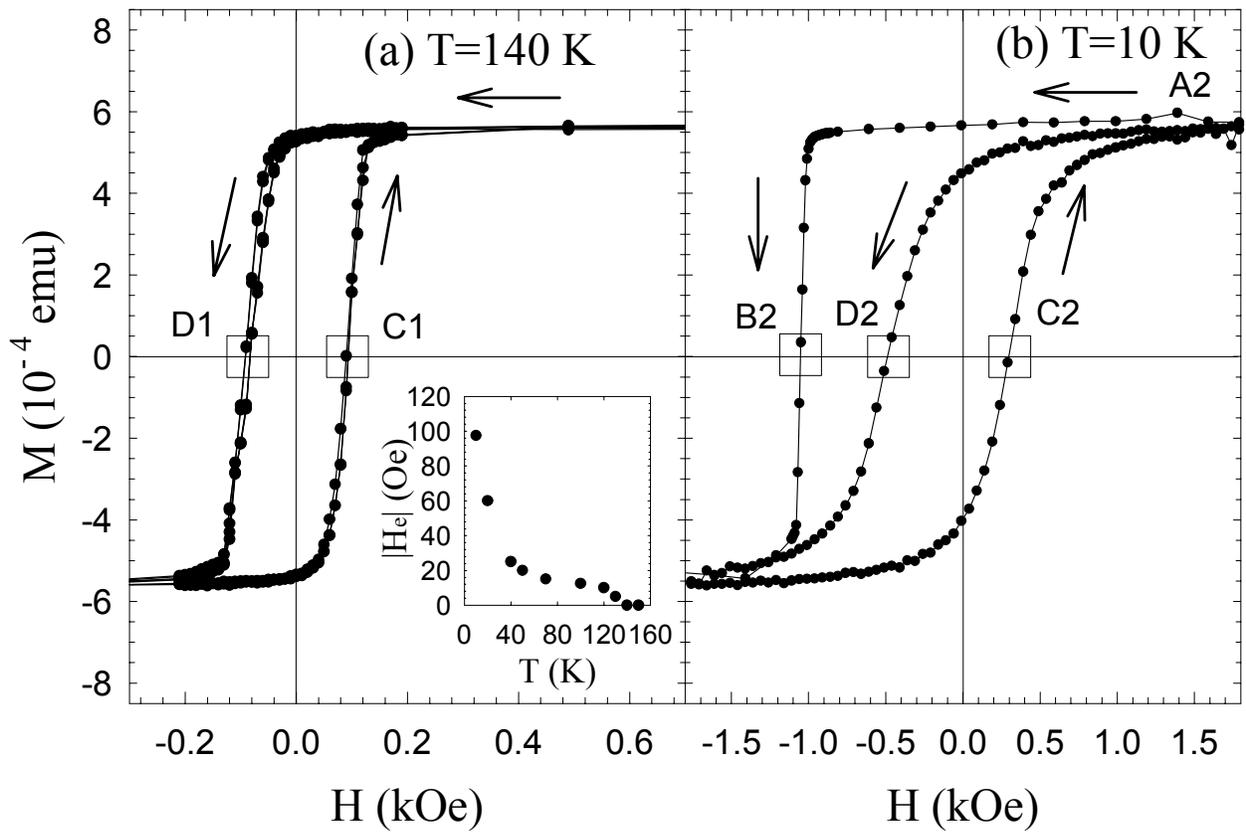

Figure 2

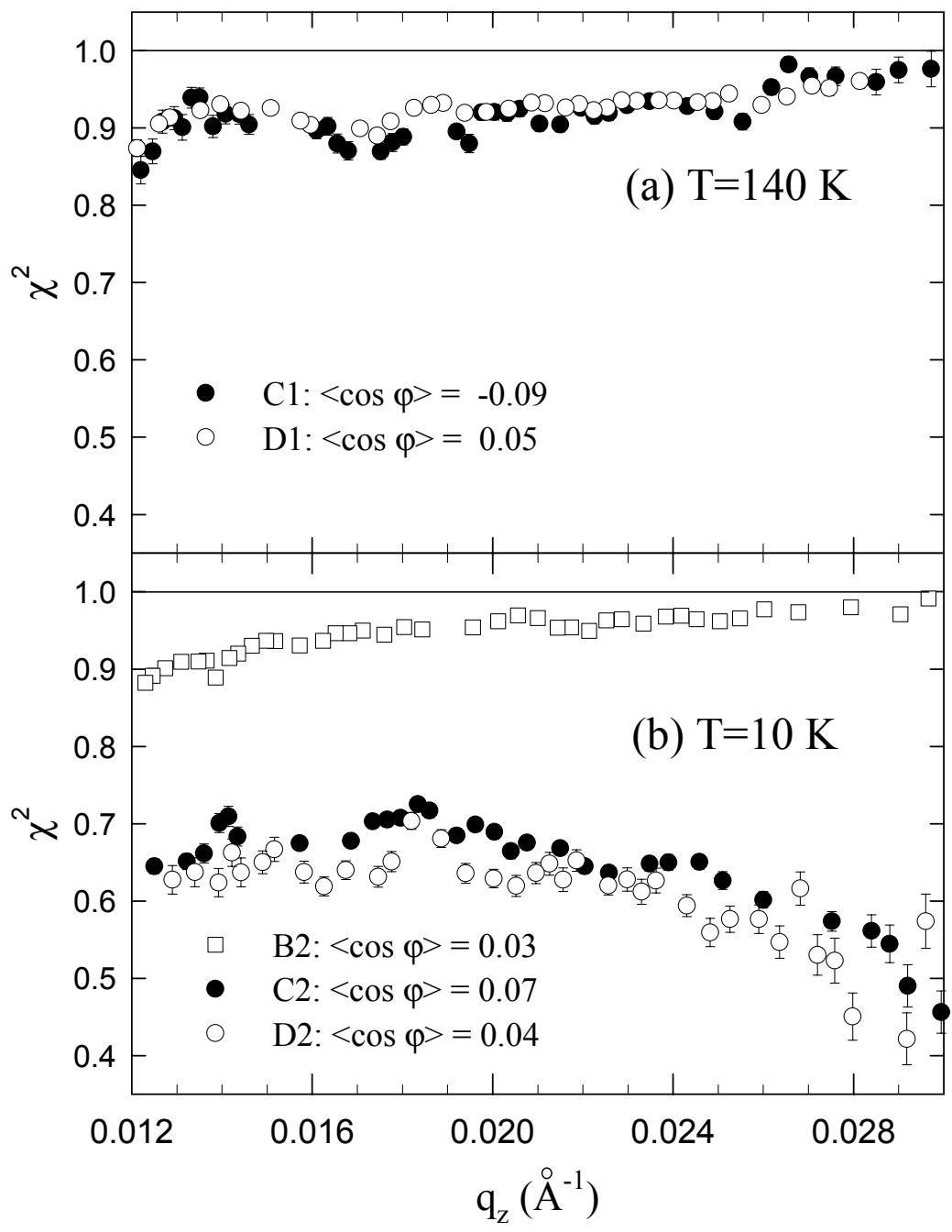

Figure 3